\documentstyle[12pt]{article}
\newcommand{\prepr}[1] {\begin{flushright} {\bf #1} \end{flushright} 
  \vskip 1.5cm} 
\newcommand{\titul}[1] {\begin{center}{\large\bf #1 } 
\end{center}\vskip 1.cm}

\newcommand{\abstr}[1] {{\begin{center} \vskip .5cm {\bf Abstract
                        \vspace{0pt}} \end{center}}\begin{quote} #1
                        \end{quote}}

\begin{document}

\begin{titlepage}
\prepr{INR-0919/96\\JINR E2-96-91\\US-FT/14-96
\\May  1996}
\titul{Next-to-next-to-leading order  QCD
analysis  of the CCFR data for  $xF_3$
and $F_2$ structure functions of the deep-inelastic neutrino-nucleon
scattering}
\begin{center}
{\bf A.L. Kataev$^{1}$,~~A.V. Kotikov$^{2,3}$,~~~G. Parente$^{4}$ and
A.V. Sidorov$^{5}$}\\ [1cm]

$^{1}$ {\em Institute for Nuclear Research of the Academy
of Sciences of Russia, 117312 Moscow, Russia} \\ $^{2}$ {\em ENSLAPP, 
 F-33175 Annecy-le Vieux, France}\\ $^{3}$ {\em Particle Physics 
 Laboratory, Joint Institute for Nuclear Research, 141980 Dubna, 
Russia}\\ $^{4}$ {\em Department of Particle Physics, University of 
Santiago de Compostela, 15706 Santiago de Compostela, Spain} \\
$^{5}$ {\em Bogoliubov Laboratory of Theoretical Physics, Joint 
Institute for Nuclear Research, 151980 Dubna, Russia}

\end{center}

\abstr{The next-to-next-to-leading order (NNLO) non-singlet 
QCD analysis of the experimental data of the CCFR collaboration 
of Ref.[1] for the $xF_3$ and $F_2$  structure functions of the 
deep-inelastic scattering of neutrinos and antineutrinos on the 
nucleon by means of the Jacobi polynomial expansion method is made. 
The target mass corrections are also taken into account. We 
demonstrate that the NNLO corrections have the important effect: 
they are decreasing  the difference between the values of 
$\alpha_s(M_Z)$, extracted from the NLO fit of $xF_3$ data and from 
the NLO non-singlet fit of the combined $xF_3$ and $F_2$ CCFR data. 
The obtained NNLO results for $\alpha_s(M_Z)$ are :
$\alpha_s(M_Z)=0.109 \pm 0.003 (stat) \pm 0.005 (syst) \pm 0.003 
(theor)$ (from $xF_3$ data) and 
$\alpha_s(M_Z)=0.111 \pm 0.002 (stat) \pm 0.003 (syst) \pm 0.003
(theor)$ (from the combined $xF_3$ and $F_2$ data).
We also estimate the values of the order $O(\alpha_s^2)$-corrections 
to the Gottfried sum rule.} 

\end{titlepage}

{\bf 1.}~The recent  measurements of the CCFR collaboration
provide the most precise up to now experimental results for the
structure functions (SFs) $xF_3(x,Q^2)$ and $F_2(x,Q^2)$ of the
deep-inelastic scattering (DIS) of neutrinos and antineutrinos
on nucleons \cite{CCFR}. The next-to-leading order (NLO) QCD analysis
of the available at present data, obtained at Fermilab Tevatron, was 
made in Ref.\cite{CCFR} with the help of the approach, based on the
solution of the corresponding  integro-differential
equation \cite{AP}, \cite{Lipatov}.  The results for the parameter 
$\Lambda_{\overline{MS}}^{(4)}$ were extracted separately from the 
fits of the observed behavior of the  $xF_3$ SF and from the 
nonsinglet (NS) analysis  of the combined available data for $xF_3$ SF 
at $x<0.5$ and for $F_2$ SF at $x>0.5$ in the energy regions where the 
higher-twist corrections were considered as the negligible small 
effects.  In the process of the latter fits the statement was made 
that  the substitution of the data for $xF_3$ by the data for $F_2$ at 
$x>0.5$ is improving the experimental precision of the results 
obtained  due to higher statistics of $F_2$ in this region.

The value of the parameter
$\Lambda_{\overline{MS}}^{(4)}$, which  resulted from
 the NLO fits of the data for $xF_3$ SF, is \cite{CCFR}:
\begin{equation}
\Lambda_{\overline{MS}}^{(4)}=171 \pm 32(stat)\pm54 (syst)~MeV
\label{1}
\end{equation}
while the combined NLO NS analysis  of $xF_3$ and $F_2$ SFs data gave
\cite{CCFR}
\begin{equation}
\Lambda_{\overline{MS}}^{(4)}=210\pm 28 (stat)\pm 41(syst)~MeV~~~.
\label{2}
\end{equation}
Let us remind  that there is also
another method of the analysis of DIS data, which is based on the
idea of the reconstruction of the SF from its Mellin moments
using the expansion of the SF into the series over the
Jacobi polynomials \cite{PS}. This method was developed in
Ref.\cite{ChR} and  Refs.\cite{Kri,Kri1} and
successfully applied in the process of the NLO fits of DIS data of the
BCDMS collaboration \cite{BCDMS} and the mentioned above data of 
the CCFR collaboration \cite{KatSid,CR}.

In the last several years the considerable
progress was achieved in the area of the analytical evaluation
of the characteristics of DIS  at  the next-to-next-to-leading order
(NNLO) of perturbative QCD in the $\overline{MS}$-scheme.  Indeed,
the NNLO corrections were found  in Ref.\cite{VZ} for the
coefficient functions of the SF $F_2$, while in Refs.\cite{VZ2,VZ3}
the similar NNLO calculations were made for the SF $xF_3$. The
result of
Ref.\cite{VZ2} is in agreement with the one for the NLO
corrections to the $n=1$ moment of $xF_3$, namely to the
Gross-Llewellyn Smith sum rule \cite{GL},
while the results
of Ref.\cite{VZ} are in agreement with the expressions
 for the NNLO corrections to  the coefficient functions of the NS
moments of $F_2$ SF, calculated analytically in the case of the even
moments with $n=2,4,6,8,10$ \cite{LV}.
Moreover, in the case of
$n=2,4,6,8$ the analytical expression of the NNLO corrections to the
anomalous dimensions of the NS moments of $F_2$ SF are  
known at present  \cite{LRV}.  
Quite recently the analytical result for the NNLO correction 
to the anomalous dimension of even $n=10$ NS moment of $F_2$ SF became 
available \cite{new}.
Together with the
expression for the NNLO
correction to the QCD $\beta$-function \cite{Lesha},
which was confirmed in Ref.\cite{LV1}, the results of
Refs.\cite{VZ}-\cite{new}  are forming the
theoretical background for the extraction of the value of the
parameter $\Lambda_{\overline{MS}}^{(4)}$ from the NS fits of
the DIS data at the level of new theoretical precision, namely with
taking into account the effects of the NNLO corrections.

We think that it is still rather difficult to 
push this program ahead  in the framework of the integro-differential 
formalism.  Indeed, the analytical dependence of the NNLO corrections 
to the anomalous dimensions from the number of the related moments $n$ 
is unknown at present. Moreover,  in view of the definite limitations 
of the possibilities of the distinguished methods of 
Ref.\cite{methods}, used in the calculations of 
Refs.\cite{GL}-\cite{new}, \cite{LV1}, \cite{Tolya}, this 
information can not be obtained without additional theoretical efforts 
\footnote{The calculating method, analogous to the ones developed in 
the first two works of Ref.\cite{methods} was also developed for other 
purposes in Ref.\cite{Vassiliev}}. This fact does not allow one to 
calculate at present explicitly the NNLO correction to the  kernel of 
the corresponding integro-differential equation.

It should be stressed that in 
the framework of the method of the Jacobi polynomials  it is necessary 
to use the information about the renormalization-group evolution only 
of the fixed number of moments. To study the evolution of the NS 
moments at the NNLO level it is possible to use the results of the 
exact calculations of Refs.\cite{VZ}-\cite{VZ3} and to supplement the 
information about the NNLO coefficients of the NS anomalous dimensions 
of the certain even moments \cite{LRV} by some smooth interpolation 
procedure \cite{PKK}. In Ref.\cite{PKK} this program was realized for 
the extraction of the value of the parameter 
$\Lambda_{\overline{MS}}^{(4)}$ from the NS  fits of the data for the 
$F_2$ SF, provided by the BCDMS collaboration, in the approximation 
when the effects of the target mass corrections (TMC) were neglected. 
Note, however, that this analysis is non-complete, since in the case 
 of the NNLO fits  of the data for the DIS of charged leptons on 
 nucleons it is also necessary to take into account the information 
 about the NNLO QCD corrections to the singlet coefficient functions 
 and singlet anomalous dimensions, recently calculated in Ref. 
\cite{new}.

This paper is devoted to the NNLO NS QCD analysis of the available  
data provided  by the CCFR collaboration \cite{CCFR}. First, in order 
to compare the outcomes of the applications of the Jacobi polynomials 
expansion method with the NLO results, obtained by the CCFR 
collaboration, we are performing the detailed NLO fits with taking 
into account the TMC \footnote{ In the 
similar analysis of Ref.\cite{KatSid,CR} the effects of the TMC were 
 neglected.}. Then we are performing the detailed NNLO fits of the 
available CCFR data.  Using the certain  
interpolation/extrapolation procedure we are also presenting the 
estimates of the previously unknown order $O(\alpha_s^2)$-corrections 
to the Gottfried sum rule.

{\bf 2.}~We first discuss the theoretical basis of the method used in this
work.  
Let us  define the Mellin moments for the NS SF $xF_3(x,Q^2)$:
$M_n^{NS}(Q^2)=\int_0^1 x^{n-1}F_3(x,Q^2)dx$
where $n=2,3,4,...$. The theoretical expression for these moments
obey the following renormalization group equation
$\bigg(\mu\frac{\partial}{\partial\mu}+\beta(A_s)\frac{\partial}{\partial A_s}
-\gamma_{NS}^{(n)}(A_s)\bigg)M_n^{NS}(Q^2/\mu^2,A_s(\mu^2))=0$
where $A_s=\alpha_s/(4\pi)$.
For the simplicity we consider the case when the factorization scale
and the scale of the ultraviolet renormalizations are equal to each other.
The renormalization group functions are defined as
$\mu\frac{\partial A_s}{\partial\mu}=\beta(A_s)=-2\sum_{i\geq 0}
\beta_i A_s^{i+2}$~~,
$\mu\frac{\partial ln
Z_n^{NS}}{\partial\mu}=\gamma_{NS}^{(n)}(A_s) =\sum_{i\geq 0}
\gamma_{NS}^{(i)}(n) A_s^{i+1}$
where
$ Z_n^{NS} $
are the renormalization constants of the corresponding
NS operators.  The solution of the renormalization-group 
equation  
can be presented in the following form \begin{equation}
\frac{M_n^{NS}(Q^2)}{M_n^{NS}(Q_0^2)}=exp\bigg[\int_{A_s(Q_0^2)}^{A_s(Q^2)}
\frac{\gamma_{NS}^{(n)}(x)}{\beta(x)}dx\bigg]\frac{C_{NS}^{(n)}(A_s(Q^2))}
{C_{NS}^{(n)}(A_s(Q_0^2))}
\label{7}
\end{equation}
where $M_n^{NS}(Q_0^2)$ is the phenomenological quantity related to the
factorization scale dependent factor.
It can be parametrized through the parton distributions at fixed momentum
transfer $Q_0^2$. In our studies we will consider one of the possible
forms of the parametrization of $xF_3$ SF, choosing
$M_n^{NS}(Q_0^2)=\int_0^1 x^{n-2} A(Q_0^2)x^{b(Q_0^2)}(1-x)^{c(Q_0^2)}
(1+\gamma(Q_0^2)x)dx$
with  $\gamma \neq 0$.

At the NNLO the  expression for
the coefficient function $C_{NS}^{(n)}$  has the 
following general form 
$C_{NS}^{(n)}(A_s)=1+C^{(1)}(n)A_s+C^{(2)}(n)A_s^2$,
while the corresponding expansion of the anomalous dimensions term in
Eq.(\ref{7}) is
\begin{equation}
exp\bigg[\int^{A_s(Q^2)}
 \frac{\gamma_{NS}^{(n)}(x)}{\beta(x)}dx\bigg]=
\big(A_s(Q^2)\big)^{\gamma_{NS}^{(0)}(n)/2\beta_0}[1+p(n)A_s(Q^2)+q(n)
(A_s(Q^2))^2]
\label{11}
\end{equation}
where $p(n)$ and $q(n)$ are defined as
$p(n) = \frac{1}{2}\left [ \frac{\gamma_{NS}^{(1)}(n)}{\beta_{1}} -
\frac{\gamma_{NS}^{(0)}(n)}{\beta_{0}}\right 
]\frac{\beta_1}{\beta_0}$,   
$q(n) = 
\frac{1}{4} [ 2 p(n)^2 +{\frac{\gamma_{NS}^{(2)}(n)}{\beta_0}}
+$ 
$\frac{(\beta_1^2-\beta_2\beta_0)}{\beta_0^3}\gamma_{NS}^{(0)}(n)
-\gamma_{NS}^{(1)}(n)\frac{\beta_1}{\beta_0^2}
 ]$ .
Notice that in our normalizations
 $\beta_0$, $\beta_1$ and $\beta_2$ read
$\beta_0=11-(2/3)f$,
$\beta_1=102-(38/3)f$,
$\beta_2=(2857/2)-(5033/18)f+(325/54)f^2$.

Now we are ready to present the basic formula for
the $Q^2$-evolution of the  NS-moments, which will be used
in our fits:
\begin{equation}
\frac{M_n^{NS}(Q^2)}{M_n^{NS}(Q_0^2)}=\bigg[\frac{A_s(Q^2)}{A_s(Q_0^2)}\bigg]
^{\gamma_{NS}^{(0)}(n)/2\beta_0}\frac{
[1+p(n)A_s(Q^2)+q(n)(A_s(Q^2))^2]}
{[1+p(n)A_s(Q_0^2)+q(n)(A_s(Q_0^2))^2]}\frac{C_{NS}^{(n)}(A_s(Q^2))}
{C_{NS}^{(n)}(A_s(Q_0^2))}
\label{mq}
\end{equation}
Eq.(\ref{mq}) can be expressed
in terms
of the inverse powers of the $\ln(Q^2/\Lambda_{\overline{MS}}^2)$
using the traditionally considered expansion
\begin{eqnarray}
A_s(Q^2)=\frac{\alpha_s(Q^2)}{4\pi}=\frac{1}{\beta_0
ln(Q^2/\Lambda_{\overline{MS}}^2)}- 
\frac{\beta_1
lnln(Q^2/\Lambda_{\overline{MS}}^2)}{\beta_0^3
ln^2(Q^2/\Lambda_{\overline{MS}}^2)} \\ \nonumber         
+\frac{\beta_1^2 (ln^2~ln(Q^2/\Lambda_{\overline{MS}}^2)-
lnln(Q^2/\Lambda_{\overline{MS}}^2)+\frac{\beta_2\beta_0}{\beta_1^2}-1)}
{\beta_0^5ln^3(Q^2/\Lambda_{\overline{MS}}^2)}
\label{alnlo}
\end{eqnarray}

The similar analysis can be also done for the NS part of
$F_2$ SF. This analysis can be applied in the kinematical region 
 $x>0.5$ where the valence quark distributions are dominating over the 
distributions of the sea quarks and gluons. Therefore in this
kinematical region  the singlet nature of $F_2$ SF can be neglected
and  the approximation $xF_3=F_2$, considered by the CCFR
collaboration \cite{CCFR}, can be  used after taking into account the
effects of the TMC. It should be noted, however, that this procedure
has definite theoretical uncertainties, since the NS coefficient
functions and the anomalous dimension terms of the $xF_3$ and $F_2$
SFs are rigorously speaking slightly different (see 
discussions below).

In order to include  the TMC  in our  fits 
we  introduce
the Nachtmann moments \cite{Nacht} of the SFs
$F_2$ and $F_3$:
\begin{eqnarray}
M_{n,F_2}^{TMC}(Q^2)=\int_{0}^{1}dx\frac{\xi^{n+1}}{x^3}F_2(x,Q^2)
\frac{(3+3(n+1)V+n(n+2)V^2)}{(n+2)(n+3)},
\label{f2}
\end{eqnarray}
\begin{eqnarray}
M_{n,xF_3}^{TMC}(Q^2)=\int_{0}^{1}dx\frac{\xi^{n+1}}{x^3}F_3(x,Q^2)
\frac{(1+(n+1)V)}{(n+2)},
\label{f3}
\end{eqnarray}
where
$\xi=2x/(1+V)$, $V=\sqrt{1+4M_{nucl}^2x^2/Q^2}$ and
$M_{nucl}$ is the mass of a nucleon.
These expressions  could be expanded into a series in
powers of $M_{nucl}^2/Q^2$ \cite{GEORGI}. Taking into account the
order $O(M_{nucl}^4/Q^4)$ corrections, we get
\begin{eqnarray}
M_{n,F_2}^{TMC}(Q^2)&=&M_n^{NS}(Q^2)+\frac{n(n-1)}{n+2}\frac{M_{nucl.}^2}{Q^2}
M_{n+2}^{NS}(Q^2)\\ \nonumber
&&+\frac{(n+2)(n+1)n(n-1)}{2(n+4)(n+3)}
\frac{M_{nucl.}^4}{Q^4}M_{n+4}^{NS}(Q^2)
+O(\frac{M_{nucl}^6}{Q^6}),
\label{m2}
\end{eqnarray}
\begin{eqnarray}
M_{n,xF_3}^{TMC}(Q^2)&=&M_n^{NS}(Q^2)+\frac{n(n+1)}{n+2}\frac{M_{nucl.}^2}{Q^2}
M_{n+2}^{NS}(Q^2) \\ \nonumber
&&+\frac{(n+2)(n+1)n}{2(n+4)}
\frac{M_{nucl.}^4}{Q^4}M_{n+4}^{NS}(Q^2)
+O(\frac{M_{nucl}^6}{Q^6}),
\label{m3}
\end{eqnarray}
We have checked that the the influence  of the order
$O(\frac{M_{nucl}^4}{Q^4})$ terms in Eqs.(9),(10)
to the outcomes  of the concrete LO and NLO fits is very small. 
Therefore, in what follows we will use only the first two terms in the 
r.h.s. of Eqs.(9),(10).

The SF is reconstructed from the corresponding Mellin moments
using the following equation
\begin{equation}
xF_{3}^{N_{max}}(x,Q^2)=x^{\alpha}(1-x)^{\beta}\sum_{n=0}^{N_{max}}
\Theta_n ^{\alpha , \beta}
(x)\sum_{j=0}^{n}c_{j}^{(n)}{(\alpha ,\beta )}
M_{j+2,xF_3}^{TMC} \left ( Q^{2}\right ),
\label{e7}
\end{equation}
where $\Theta_n^{\alpha,\beta}$ are the Jacobi polynomials and 
$\alpha,\beta$ are their parameters, fixed by the condition of the 
requirement of the minimization of the error of the reconstruction of 
the SF (see Refs.\cite{Kri,Kri1}).  In our analysis
 we are considering the region $6\leq N_{max}\leq 10$, which
corresponds to the number of moments  $2\leq n\leq N_{max}+2$, or to 
$2\leq n\leq N_{max}+4$ if the TMC are included up to   order 
 $O(M_{nucl}^2/Q^2)$-terms in the r.h.s. of Eqs.(9),(10).   Thus we 
will be interested in the study of the behavior of the moments with 
$2\leq n\leq 14$.

{\bf 3.}~~~Let us now summarize the available perturbative QCD
 information for the coefficient functions  of the NS Mellin moments
in the $\overline{MS}$-scheme. The corresponding NLO
coefficients of the Mellin moments of the NS SF $F_2$ was calculated
in Ref.\cite{Bardeen}. Its exact analytical expression reads
$C_{F_2}^{(1)}(n)=\frac{4}{3}\bigg[2[S_1(n)]^2+3S_1(n)-2S_2(n)-\frac{2S_1(n)}
{n(n+1)}
+\frac{3}{n}+\frac{4}{n+1}+\frac{2}{n^2}-9\bigg]$
where
$S_k(n)=\sum_{j=1}^{n}\frac{1}{j^k}$.
The similar term in the expression for the coefficient function
of the Mellin moments of the SF
$xF_3$ receives the following additional contribution 
$C_{F_3}^{(1)}(n)=C_{F_2}^{(1)}(n)-
\frac{8}{3}\frac{2n+1}{n(n+1)}$ \cite{Bardeen}.
Since we will be interested in the numerical values of
the corresponding coefficients of the coefficient function
$C_{NS}^{(n)}(A_s)$ for the fixed number of the moments, we present
them in the numerical form in Table 1, choosing the number of 
flavors $f=4$. The results for the NNLO coefficients 
$C_{F_2}^{(2)}(n)$ and $C_{F_3}^{(2)}(n)$ are obtained from the 
results of Refs.[11,12] and Ref.[13] respectively taking the 
corresponding Mellin moments \footnote{The term proportional to
$C_AC_Fln(1-z)$ in Eq.(15) of Ref.[12] contains the misprint. It 
should be changed to $-(50/3-z142/3)$. After this change the results 
for the related moments coincide with the results given in Table 2 of 
Ref.[13].}. 

Notice, that the factor
$\Delta_C^{(1)}(n)= C_{F_2}^{(1)}(n)-C_{F_3}^{(1)}(n)$ is decreasing
with the increase of the number of moment. The similar feature also
holds at the NNLO level for the difference
$\Delta_C^{(2)}(n)=C_{F_2}^{(2)}(n)-C_{F_3}^{(2)}(n)$ (see Table 1).

We will discuss now the contributions to the NS anomalous dimensions
of the Mellin moments of $F_2$ and $xF_3$ SFs. The expressions for 
the leading order coefficients
$\gamma_{NS,F_2}^{(0)}(n)$ and $\gamma_{NS,F_3}^{(0)}(n)$ coincide
with each other and read
$\gamma_{NS}^{(0)}(n)=\frac{8}{3}\bigg[4S_1(n)-\frac{2}{n(n+1)}-3\bigg]$.
The NLO coefficients of the NS anomalous dimensions were first
obtained in Ref.\cite{FRS}. These results were presented later on in
the more simplified form \cite{Pako}. 
The NS results of Refs.\cite{FRS,Pako} were confirmed by the 
independent calculation in Ref.\cite{CFP}. It should be stressed, that 
starting from the NLO the analytical expressions of anomalous 
dimensions of the NS moments of $F_2(xF_3)$ SFs can be obtained from 
the direct calculations of the corresponding Feynman diagrams in 
the momentum space only in the concrete cases of the even (odd) number 
of the Mellin moments. To find out the values of these anomalous 
dimensions at any $n$ the procedure of the analytical continuation 
should be performed for the cases of odd (even) number of moments 
\footnote{In principle, the corresponding calculations in the 
x-space lead to the complete expressions for the anomalous 
dimensions without application of the procedure of the analytical 
continuation.}. This procedure can be realized using the results of 
the considerations of the first work of Ref.\cite{Tolya} and of 
Ref.\cite{Kri1} (the related analysis was described in more detail 
in Ref.\cite{KL}).  
In the cases of $F_2(xF_3)$ SFs the 
procedure results in the application of the ``+" (``$-$") 
prescriptions of Ref.\cite{Kri1}, which shift slightly up (down) the 
values of the corresponding anomalous dimensions of the odd (even) 
moments.

We have noticed, as the authors of Ref.\cite{Kri1}, that even though
the applications of the prescriptions mentioned above produce a very
small change of the values of the NLO corrections to the NS anomalous
dimensions, they are extremely important for the reconstruction of
the SFs using the Jacobi polynomial method. In fact when high moments
with $n>10$ are involved in the polynomial reconstruction, the
existence of very large factorial terms in the Jacobi polynomial
expansion  spoil the convergence of the fitting procedure if the
effects of the analytical continuation are not taken into account.

To save the space we will not present here the final analytical
expressions for $\gamma_{NS,F_2}^{(1)}(n)$ and
$\gamma_{NS,F_3}^{(1)}(n)$ through the poligamma-functions and some
other associated to them functions, which are continuous in its
argument $n$. However, in Table 2 we present their numerical values
for $1\leq n\leq 14$, fixing the number of flavors $f=4$. We
emphasize, that using the $1/n$-expansion (which turns out to be the
good approximation starting already from $n=4$) it is possible to
show, that the difference between the final analytical expressions of
$\gamma_{NS,F_2}^{(1)}(n)$ and $\gamma_{NS,F_3}^{(1)}(n)$ is rather
small, namely
$\Delta_{\gamma}(n)=\gamma_{NS,F_3}^{(1)}(n)-\gamma_{NS,F_2}^{(1)}(n)
\approx 128/(3n^6)+O\bigg(1/n^7\bigg)$ \cite{RS}.
For $n=2,4,6,8$ the concrete numerical values of
$\Delta_{\gamma}(n)$ are listed in Table 3. For larger even values of
$n$ the non-rounded off values of $\Delta_{\gamma}$ are less than
$10^{-4}$.

Assuming that the feature, similar to discussed above,
will remain true at the NNLO also, we will use the calculated
in Ref.\cite{LRV,new} values of $\gamma_{NS,F_2}^{(2)}(n)$ in the NNLO
analysis of $xF_3$ data. In analogy with the considerations of 
Ref.\cite{PKK}, the numerical values of
$\gamma_{NS,F_2}^{(2)}(n)\approx\gamma_{NS,F_3}^{(2)}(n)$ for
$n=3,5,7,9$ has been obtained from the results of calculations of
Ref.\cite{LRV,new} (made for the case of $n=2,4,6,8$ in 
Ref.\cite{LRV} and of $n=10$ in Ref.\cite{new}) using the smooth 
interpolation. 
The concrete numerical expressions for the coefficients
$\gamma_{NS}^{(2)}(n)$  are presented in Table 2
for $2\leq n\leq 10$.
It should be stressed, that in order to verify the validity   
of the interpolation procedure used by us, it is of real 
interest to calculate the values of $\gamma_{NS,F_3}^{(2)}(n)$ 
analytically in the case of $n=3,5,7,9$. 
Moreover, it is also rather desirable to learn the explicit expression 
for this anomalous dimension in the case of $n=11$.  In principle, 
these calculations can be done with the help of the analytical methods 
of Ref.\cite{methods}. 

In order to get rather rough estimate of the 
value of $\gamma_{NS,F_2}^{(2)}(1)$ we extrapolated the smooth curve, 
obtained by the interpolation  procedures described above, to the 
point $n=1$ and obtained $\gamma_{NS,F_2}^{(2)}(1)=306.7$ for $f=4$.
Notice, that while the exact expressions for $\gamma_{NS}^{(0)}=0$ and 
$\gamma_{NS,F_2}^{(1)}(1)=2.5576$ do not depend from the number of 
flavors $f$ taken into account, the value $\gamma_{NS,F_2}^{(2)}(1)$ 
can be $f$-dependent. Indeed, repeating our  
interpolation/extrapolation procedure  in the case of $f=3$ we get 
with the non-controllable error-bars. Using this value as the input 
parameter in the renormalization-group expression for the first NS 
moment of $F_2$ SF we arrive to the following order $O(\alpha_s^2)$ 
estimated expression for the Gottfried sum rule in the case of 
$\overline{u}-\overline{d}$ flavor-symmetrical sea for $f=3$ numbers 
of active flavors:  \begin{equation}
GSR(Q^2)=\int_0^1[F_2^{lp}(x,Q^2)-F_2^{ln}(x,Q^2)]\frac{dx}{x}=
\frac{1}{3}\bigg[1+0.036\bigg(\frac{\overline{\alpha}_s}{\pi}\bigg)
+0.72\bigg(\frac{\overline{\alpha}_s}{\pi}\bigg)^2\bigg]
\label{Got}
\end{equation}
where $\overline{\alpha}_s=\alpha_s(Q^2)$.
In the case of $f=4$ numbers of flavours the analogous expression 
reads $GSR(Q^2)=(1/3)\bigg[1+0.038(\overline{\alpha}_s/\pi)
+0.55(\overline{\alpha}_s/\pi)^2\bigg]$.
Thus, taking into account the small {\bf positive} QCD corrections to
this sum rule, including 
the newly estimated order $O(\alpha_s^2)$-contribution, can not
explain the deviation of the theoretical prediction of 
Eq.(\ref{Got}) from the experimental result of the NMC collaboration, 
namely $GSR=0.235\pm 0.026$ at $Q^2=4~GeV^2$ \cite{NMC} without the 
assumption, that the light quark sea is flavor asymmetric, which means 
that $\overline{u}-\overline{d}>0$.

{\bf 4.}~~~The concrete fits of the available CCFR data of 
Ref.\cite{CCFR}, made by means of the Jacobi polynomial expansion 
method, were performed by us for different values of the  
scale $Q_0^2$  with the 
help of two independently written computer programs. 
The fisrt Program  has been created as the result of the works of 
Refs.\cite{Kri,Kri1}, while the second Program  previously found its 
most distinguished application in the process of the fits of 
Ref.\cite{PKK}. We have checked that both Programs are giving the 
identical results and thus are compatible with each other.

In fact we confirmed the findings of Refs.\cite{Kri,Kri1} that the
results of the NS fits (namely the output values of
$\Lambda_{\overline{MS}}^{(4)}$ and of the parton distributions 
$A(Q_0^2), b(Q_0^2),c(Q_0^2),\gamma(Q_0^2)$) are non-sensitive to the 
choice of $\alpha,\beta$ provided large enough number of the Jacobi
polynomial $N_{max}$ is taken into account (we are taking 
$N_{max}=10$).  However, if smaller number of the Jacobi polynomial 
are considered, the reasonable convergence is achieved in the case of 
taking $\alpha\approx 0.7$ and $\beta\approx 3$. These values are 
allowing to approximate {\it a posteriori} with maximal precision the 
$x$-behavioral of the shape of the SF $xF_3$ by the weight function of 
the Jacobi polynomial in Eq.(11) in the scaling limit and are 
determined by the values of the parameters $b$ and $c$ 
in  the used model of the parton distributions  
(from the results of the previous Jacobi polynomial expansion fits of 
the CCFR data  one can see that $b(Q_0^2=10~GeV^2)=0.717\pm 0.012$ and 
$c(Q_0^2=10~GeV^2)=3.395\pm 0.043$ \cite{KS2}).
We faced this problem of fixing sharply the values of $\alpha,\beta$ 
in the process of the NNLO fits, which we were not able to complete 
with the reasonable precision for $N_{max}>6$ in view of the  
uncertainties of the  
extrapolation for $\gamma_{NS}^{(2)}(n)$  at $n=N_{max}+4>10$.
In principle, the question of the extension of our NNLO analysis to 
the case of $N_{max}=7$ can be studied in future  if the exact 
expressions for $\gamma_{NS,xF_3}^{(2)}(11)$  become available.

The results of the NLO fits are
summarized in Table 4, where they are compared with the ones, obtained
by the CCFR collaboration (see Ref.\cite{CCFR}). We have checked, that
the results of our  fits of the CCFR  data 
are almost nonsensitive to the variation of $Q_0^2$ in 
the region $3~GeV^2\leq Q_0^2\leq 50~GeV^2$. For
definiteness, the results of Table 4 are presented in the
case of taking $Q_0^2=10~GeV^2$.
Notice, that in spite of the fact that the methods applied in the
process of our fits are different from the ones, used by the CCFR
collaboration, the agreement with the results of Ref.\cite{CCFR} is
rather good. This feature supports further concrete applications of
the Jacobi polynomial expansion method.

Another important observation comes from the comparison of the values
for $\Lambda_{\overline{MS}}^{(4)}$, presented in Table 4, with the
outcomes of the previous fits of the available CCFR data for $xF_3$ SF 
by means of the Jacobi polynomial expansion method in the 
approximation when the TMC were neglected \cite{KatSid}. This 
comparison indicates, that the inclusion of the TMC into the analysis 
lowers the obtained NLO values of $\Lambda_{\overline{MS}}^{(4)}$ by 
over $25~MeV$. This observation confirms the previous findings of the 
IHEP-ITEP collaboration \cite{IHEP} that it is extremely important to 
take into account the effects of the TMC in the analysis of the 
neutrino-nucleon DIS data.

The inclusion of the NNLO QCD corrections into the analysis of the
available CCFR data for $xF_3$ SF made by us 
following the discussions presented above, result in the less 
vivid effect. In Table 5 we present the results of these 
fits for the value of $Q_0^2=10~GeV^2$.  One can see that for the 
maximal number of the Jacobi polynomials $N_{max}=6$, for which the 
reasonable convergence of Eq.(11) was achieved, the resulting change
$(\Delta\Lambda_{\overline{MS}}^{(4)})_{NNLO}=(\Lambda_{\overline{MS}}^{(4)})
_{NNLO}-(\Lambda_{\overline{MS}}^{(4)})_{NLO}\approx 
+2~MeV$ is very small.

The results of the combined NS fits of $xF_3$ and $F_2$ data, which 
were obtained  in the case of the cut $Q^2>15~GeV^2$, 
considered in the works of the CCFR collaboration \cite{CCFR}, are 
presented at Table 6 in the case of taking $Q_0^2=10~GeV^2$. 
It should be  noted, that the combined analysis of $xF_3(x<0.5)$
and $F_2(x>0.5)$ CCFR data of Ref.\cite{CCFR} was made  
using the following approximate conditions: first the TMC in 
Eqs.(9), (10) were taken into account separately to the NS 
moments $M_{n,F_3}^{TMC}(Q^2)$ and $M_{n,F_2}^{TMC}(Q^2)$, but then 
both remaining moments $M_{n,F_3}^{NS}(Q^2)$ and $M_{n,F_2}^{NS}(Q^2)$ 
were considered to be equal to each other, parametrized at the scale
$Q_0^2$ by the same way and evolved to the points $Q^2$
(where the experimental data is available) using the same
renormalization-group equations with the same
coefficient functions and anomalous dimensions as in the case of the
renormalization group evolution of the Mellin moments of $xF_3$ SFs
(see Table 1 and Table 2). Then the Jacobi polynomial expansion of
Eq.(8) was applied in the regions $x<0.5$ and $x>0.5$ with the
same parameters $\alpha$ and $\beta$. This approximate procedure has
some theoretical support. Indeed, it can be shown, that in the region
of large enough values of $x$, namely $0.5<x<1$, the behavior of the
reconstructed NS SF in Eq.(11) is mainly determined by the Mellin
moments $M_n^{NS}(Q^2)$ with large enough $n$, for which the values
of the factors
$\Delta_C^{(1)}(n)/4=(C_{F_2}^{(1)}(n)-C_{F_3}^{(1)}(n))/4$,
$\Delta_C^{(2)}(n)/16=(C_{F_2}^{(2)}(n)-C_{F_3}^{(2)}(n))/16$ and
$\Delta_{\gamma}(n)=\gamma_{NS,F_3}^{(1)}(n)-\gamma_{NS,F_2}^{(1)}(n)$
are rather small. The applied approximation allows us not to 
double the number of the parameters of the fits, namely the Jacobi 
polynomials parameters $\alpha,\beta$ and the parameters of the parton
distributions $A,b,c$ and $\gamma$.

However, in view of the violation of the approximations 
$\Delta_C^{(1)}(n)=0$, $\Delta_C^{(2)}(n)=0$, $\Delta_{\gamma}(n)=0$ 
for the moderate values of $n$ (see Table 1 and Table 2),   
this approximate procedure has definite 
theoretical uncertainties. Thus, to our point of view, the price for 
the decreasing the experimental uncertainties due to the inclusion in 
our  fits of the  $F_2$ CCFR data in the region of $x>0.5$, which has 
higher statistical maintenance, is the increase of the theoretical 
uncertainties. The similar problem is also appearing in the course 
of the combined NLO fits of the CCFR data made in Ref.\cite{CCFR}  
using the 
approximation $xF_3=F_2$ at $x>0.5$. Therefore, we think that in the 
summary tables of $\alpha_s$ measurements it is better to present not 
only the results of the fits of the combined $xF_3$ and $F_2$ CCFR 
data, like it was done say in Ref.\cite{Rev}, but the results of the 
separate fits of $xF_3$ CCFR data also.

{\bf 5.}~~~
In order to study the 
structure of the perturbative QCD predictions for the NS moments
of the SFs of the DIS neutrino-nucleon scattering, we
performed the fits of the available CCFR data of Ref.\cite{CCFR}
at the LO also. The
value for $\Lambda_{LO}^{(4)}$ turned out to be very stable to the
variation of $Q_0$. For the cut $Q^2>10~GeV^2$ the result is
$\Lambda_{LO}^{(4)}=131\pm 26~MeV$ in the case of
the analysis of the $xF_3$ data, while in the case of the combined  
analysis of the $xF_3$ and $F_2$ data with the cut $Q^2>15~GeV^2$ we 
got   $\Lambda_{LO}^{(4)}=208\pm 26~ MeV$. 
Comparing these results with 
other ones presented in Tables 4-6  we arrive to the 
following conclusions \begin{itemize}
\item the effects of the NLO QCD corrections are important and are 
increasing the values of the QCD scale parameter by over $50~MeV$
in the case of $xF_3$ fit and by over $25~MeV$ in the case of the 
combined NS fits of $xF_3$ and $F_2$ data;
\item on the contrary to the 
findings revealed in the process of the NNLO NS analysis of the BCDMS 
data for $F_2$ SF of the muon-nucleon DIS, which was made in 
Ref.\cite{PKK}, the inclusion of the NNLO QCD corrections 
into the analysis of the available CCFR data 
for $xF_3$ SF do not affect essentially  
the values of $\Lambda_{\overline{MS}}^{(4)}$, extracted at 
the NLO level; 
\item this 
effect is typical to the analysis of the data for $xF_3$ 
SF only. Indeed, after  taking into account the CCFR 
data for $F_2$ we found that the inclusion of the NNLO 
QCD corrections into the analysis leads to the decrease of the   
value of 
$\Lambda_{\overline{MS}}^{(4)}$
by over $25~MeV$, like it was in the case of the NNLO NS 
fits of the BCDMS data for $F_2$ SF \cite{PKK}; 
\item 
we reveal the {\bf very interesting effect} of the inclusion of the 
NNLO QCD corrections in the analysis of the CCFR data, namely the {\bf 
decrease} of the difference between the central values of the 
parameter $\Lambda_{\overline{MS}}^{(4)}$, which was extracted at the 
NLO from the separate NS fits of the $xF_3$ and $xF_3+F_2$ CCFR data 
of Ref.[1]. Indeed, at the NLO the difference was over 
$\Delta_{\Lambda_{\overline{MS}}}^{NLO}= 
(\Lambda_{\overline{MS}}^{(4)})_{xF_3} 
-(\Lambda_{\overline{MS}}^{(4)})_{xF_3+F_2}=-53~MeV$, while at the 
NNLO it is shrunk by over $50\%$ and reads 
$\Delta_{\Lambda_{\overline{MS}}}^{NNLO}=-26~MeV$ (!!);  
\item
another {\bf important effect} of the inclusion of the NNLO 
corrections was revealed as the result of the fits of the available 
$xF_3$ CCFR data with the different $Q^2$-cuts. We have found, that 
the difference between the values of $\Lambda_{\overline{MS}}^{(4)}$, 
obtained during the LO fits of the data with the cuts of 
$Q^2>15~GeV^2$ and $Q^2>5~GeV^2$, which is over $28~MeV$, is shrunk 
to $13~MeV$ in the NLO and to over $4~MeV$ at the NNLO (!!!). Thus we 
think, that in order to make the outcomes of the concrete fits less 
sensitive to the typical scales of the cuts of the experimental data 
it is very important to take into account NNLO QCD corrections.   
\end{itemize}
In order to extract from our analysis the related values of the QCD
coupling constant $\alpha_s(M_Z)$ we first evolved the results 
presented
in Table 5 and Table 6 through the threshold of the production of the
$b$-quark directly in the $\overline{MS}$-scheme using the derived in
Ref.\cite{Marciano} relation
$\Lambda_{\overline{MS}}^{(5)}=\Lambda_{\overline{MS}}^{(4)}\bigg(
\Lambda_{\overline{MS}}^{(4)}/m_b\bigg)^{2/23}ln[(m_b/
\Lambda_{\overline{MS}}^{(4)})^2]^{(-963/13225)}$~,
where we chose  $m_b\approx 4.6~GeV$.
After this we substituted the numerical expressions for
$\Lambda_{\overline{MS}}^{(5)}$ into the NLO and NNLO inverse-log
approximations for $\alpha_s$ (taking $f=5$ numbers of flavors and 
$Q^2=M_Z^2$) and obtained the results presented in Table 7, where the 
systematical uncertainties were fixed using the estimates of the 
related errors in the published results of the CCFR collaboration 
\cite{CCFR} (see Eqs.(1),(2)).

Looking carefully  at the results of Table 7 we arrive to the following
conclusions:
\begin{itemize}
\item The previously omitted from the summary tables of 
Ref.\cite{Rev} value of $\alpha_s(M_Z)$, extracted from the NLO 
analysis of the available $xF_3$ CCFR data, is in good agreement with 
the small value of $\alpha_s(M_Z)\approx 0.109$, extracted in 
Ref.\cite{Voloshin} from the QCD sum rules for the production cross 
section of bottomium states in $e^+e^-$-annihilation in the NLO of 
perturbative QCD. However, on the contrary to our analysis, the result 
of Ref.\cite{Voloshin} does not include the estimates of the possible 
effects of the NNLO perturbative QCD corrections. We think, that these 
effects can be important.  
\item Indeed, we observed, that the 
inclusion of the NNLO corrections into game  
makes the results of the fits of the available CCFR data more stable 
both to the procedure of neglecting or taking into account the data 
for $F_2$ SF at large $x$ and to the typical scales 
of the cuts of the low $Q^2$ $xF_3$ data.  
\item Our final NNLO 
results are very closed to the result 
$\alpha_s(M_Z)=0.113\pm0.003(exp)\pm0.004(theor)$ \cite{MV}, which 
comes from the NLO analysis of the BCDMS data by means of the 
integro-differential approach. In view of the appearance of 
the results of the NNLO calculations for the anomalous 
dimensions of the $F_2$ SF in the singlet case \cite{new} it 
will be of interest to refine the NS NNLO Jacobi polynomial expansion 
analysis of the BCDMS data \cite{PKK} in order to get the NNLO variant 
of the result of Ref.\cite{MV}.  \end{itemize}       

{\bf 6.}~~~The discussions of the important questions of the 
theoretical uncertainties  are now in
order.

It should be mentioned, first, that instead of using the inverse log
expansion for $\overline{\alpha}_s$ (see Eq.(6)) one can solve the 
renormalization group equation for the QCD $\beta$-function 
explicitly and arrive to the corresponding transcendental equation, 
used in Ref.\cite{PKK}. The application of this equation gives NLO 
analogs of the values of $\Lambda_{\overline{MS}}^{(4)}$, presented in 
Tables 5 and 6, systematically over $10~MeV$ smaller.
However, we have checked that this difference is compensated by the 
differences, which appear in the process of the determination of the 
values of $\alpha_s$ using the inverse logarithmic expression  and the 
solution of the presented above transcendental equation. Thus, the 
final values of $\alpha_s(M_Z)$
turn out to be almost identical.

The more important uncertainty comes from the arbitrariness in the 
choice of the matching point in the relation of Ref.\cite{Marciano}. 
(for the recent discussion of this problem see Ref.\cite{ChN}).
It was 
mentioned in the detailed reviews of Ref.\cite{Altarelli} that by 
varying the value of this point in the region $Q\approx(0.75-2.5)m_b$ 
one can estimate roughly an error of taking into account threshold 
effects. This estimate gives $\Delta\alpha_s(M_Z)=\pm 0.0015$ 
\cite{Altarelli}.  Quite recently another interesting  procedure of 
taking into account threshold effects was proposed \cite{Shirkov}. One 
can hope, that the continuing at present studies of the possibility of 
the application of this procedure in the NLO analysis of the CCFR data 
of Ref.\cite{CCFR} for the $xF_3$ SF (see Ref.\cite{MSS}) can give the 
more quantitative estimate of this theoretical uncertainty.

Another interesting problem is related to the subject of taking into 
account the nuclear effects in the analysis of the data for the
DIS of neutrinos on the targets with the fixed nuclear
content. We should warn, however, that the NLO analysis of 
Ref.\cite{ST} demonstrated that the effect of the influence of the
nuclear corrections to the procedure of the extraction of the NLO 
value of $\Lambda_{\overline{MS}}^{(4)}$ from the CCFR data of 
Ref.\cite{CCFR} for the $xF_3^{Fe}$ SF can not be large and 
is decreasing $\Lambda_{\overline{MS}}^{(4)}$ by $\leq7~MeV$ 
\cite{ST}.

As follows from the discussions presented above, the effects of the 
NNLO perturbative QCD corrections are more vivid. Apart of the 
decreasing the difference between the values of $\alpha_s(M_Z)$, 
extracted from the NLO NS fits of $xF_3$ and $xF_3+F_2$ data of 
Ref.\cite{CCFR}, the procedure of taking into account NNLO 
perturbative QCD corrections should decrease the theoretical 
uncertainty due to the arbitrariness in  the choice of the
renormalization scheme.  This effect manifested itself most obviously 
in the process of the first detailed analysis of the previous CCFR 
data for the Gross-Llewellyn Smith sum rule \cite{GLS}, which was made 
in Ref.\cite{ChK}. The effect of the reduction of the 
scheme-dependence uncertainty at the NNLO was also confirmed in the 
process of the NS analysis of the BCDMS data \cite{PKK}. We think, 
that in our NNLO results of Table 7 this uncertainty can  
change the value of $\alpha_s(M_Z)$ by over 
$\Delta\alpha_s(M_Z)\approx \pm 0.0005$, like it was in the case of 
the considerations of Ref.\cite{ChK}.

The value of this uncertainty can be simulated in part by taking into 
account the estimates of the non-calculable $N^3LO$-terms using the 
scheme-invariant methods, developed in the process of the related QCD 
studies for the DIS sum rules \cite{KatSt}. It is known from the 
considerations of Ref.\cite{SEK}, that the results of applications of 
these methods are in agreement with the analogous results, provided by 
the Pad\'e-resummation technique, which was previously used in
the process of the fits of the DIS data of the BCDMS collaboration in 
Ref.\cite{SidP}. Therefore, it might be of interest to generalize the 
studies of the work of Ref.\cite{SidP} to the case of the NNLO 
analysis of the CCFR data and thus to put the proposed above estimate 
of the scheme-dependence uncertainty of the NNLO value of 
$\alpha_s(M_Z)$ on the more solid ground.

There are also some other yet non-fixed in our studies theoretical 
uncertainties, which  come from the non-considered by us effects of
the higher-twist corrections (which, however, in the considered 
kinematical region are expected to be less important than the included 
in our analysis effects of the TMC\footnote{This statement finds its 
empirical support in the results of the NLO QCD analysis of the 
neutrino-nucleon DIS data of the CDHS and BEBC-Gargamelle 
collaborations, which was made in Ref.\cite{ALY} some time ago.})
and the
effects due to non-zero masses of the c- and b-quarks. In view of 
this we prefer to follow the point of view of Ref.\cite{Altarelli} 
that it is better to be more careful (than overoptimistic) in the 
process of estimating the total theoretical uncertainties of the 
values of $\alpha_s(M_Z)$. Thus we are presenting the following NNLO 
results for $\alpha_s(M_Z)$, extracted from the available 
CCFR data of Ref.\cite{CCFR} by means of the Jacobi polynomial 
technique:  
\begin{equation} xF_3:~~\alpha_s(M_Z)=0.109 \pm 
0.003~(stat) \pm 0.005(syst) \pm 0.003~(theor)~,
\label{alp1} \end{equation}
\begin{equation} xF_3+F_2:~~\alpha_s(M_Z)=0.111 \pm 0.002~(stat) 
\pm 0.003~ (syst) \pm 0.003~(theor)~.  \label{alp2} \end{equation}
The presented theoretical errors are including all discussed above 
uncertainties. We are also aware, that the members of the CCFR group 
are planning to update their data of Ref.\cite{CCFR}, 
treating  better the radiative EW corrections, 
dimuon and charmed mass corrections, along with extending kinematical 
coverage and reducing systematical errors \cite{Mike}. 
These studies can affect the values of the available CCFR data 
at low $x$ region (the necessity of this improvement was previously 
emphasized in Ref.\cite{parton}, devoted to the extraction of the 
parameters of the CTEQ and MRS parton distributions). In spite 
of the fact that in our analysis we have cutten down the low $Q^2$
and thus low $x$ points, it can be of real interest to perform 
the more detailed  
NNLO analysis of the updated CCFR data in future, after 
they will become available.

Another interesting problem, which is more closely related to the
extraction of the value of the Gross-Llewellyn Smith sum rule from 
the extrapolated data of the CCFR collaboration \cite{KatSid,KS2}, is
the determination of the values of the parameters of the 
parton distributions $A(Q_0^2), b(Q_0^2), c(Q_0^2), \gamma(Q_0^2)$ for 
the SF $xF_3$ from  the LO, NLO and NNLO fits.
In view of the lack of space, we are postponing the
presentation of the concrete results of our fits up to another
publication. Note, however, that 
in spite of the existence of the results of the recent 
detailed studies  of Ref.\cite{BV} 
we are still doubting whether 
this simple parametrization can be applied with the necessary
theoretical precision in the region of very small $x$. We hope, that 
the possible future theoretical considerations of the description of 
the behavior of the NS structure functions at very small $x$, which 
might be done, say, using the developed in Ref.\cite{Larry} methods, 
can be useful to clarify this interesting problem.

{\bf 7.}~~~In conclusion, we would like to stress again, that using 
the Jacobi polynomial expansion method, developed in 
Refs.\cite{ChR,Kri,Kri1}, we have shown that the effects of the NNLO 
corrections are important in the analysis of the most precise up to
now experimental data of the CCFR collaboration \cite{CCFR} for  the 
NS SFs of the neutrino-nucleon DIS.  The  obtained NNLO values of 
$\alpha_s(M_Z)$, extracted separately from the fit of the data for the 
$xF_3$ SF and from the  combined NS fit of $xF_3+F_2$ data, turn out 
to be in good agreement with the previous NLO determination of 
$\alpha_s(M_Z)$ from the fit of the BCDMS data for $F_2$ SF of the  
muon-nucleon DIS \cite{MV}. Thus the problem of understanding of the 
existing differences between the values of $\alpha_s(M_Z)$ extracted 
from DIS and from the {\bf combined} fits of the data of 4 LEP groups 
for the $Z^0\rightarrow {hadrons}$ decay width is still on the agenda.  
For the certain attempts to understand this discrepancy by introducing 
the effects of virtual SUSY particles and other effects of the 
possible new physics (see e.g. Ref.\cite{KSM} and  
Ref.\cite{Shifman}). Note, however, that there are certain  
systematical discrepancies between the values of $\alpha_s(M_Z)$, 
extracted from the $Z^0\rightarrow{hadrons}$ decay width by the 
DELPHI, ALEPH, OPAL and L3 groups separately (see e.g. \cite{Aachen}). 
Therefore, it might be also of interest to understand better the 
origin of the possible existing discrepancies.

{\bf Acknowledgements}

We are grateful to  M.H.~Shaevitz and W.G.~Seligman for 
providing us the available CCFR data of Ref.\cite{CCFR}.

During this work we gained a lot from very helpful discussions 
with G.~Altarelli, J.~Ellis, S.V.~Mikhailov, D.V.~Shirkov and A.~Vogt, 
whom we would like to thank.

A.L.K. and A.V.K. wishes to thank P.~Aurenche and P.~Sorba for 
organizing their short intersection in ENSLAPP in July 1995, which 
was very useful for the  continuation of this research.

G.P. is grateful to the organizers and to the participants of the 
Quarks-96 Seminar (Yaroslavl, May 1996), where he 
presented the part of this work, and to members of the Bogoliubov 
Laboratory of Theoretical Physics of JINR for hospitality in Dubna 
during the completing of this work.  

This work is supported by the Russian Fund for Fundamental Research, 
Grant N 96-02-18897. The work of two of us (A.L.K. and A.V.S.) 
was done within the scientific program of the INTAS project N 
93-1180. The work of G.P. was supported by CICYT and Xunta 
de Galicia ( Galician 
Autonomous Government, Spain).

\newpage

\begin{center} \begin{tabular}{||r|c|c|c|c||} \hline n&
$C_{F_2}^{(1)}(n)$ & $C_{F3}^{(1)}(n)$ & $C_{F_2}^{(2)}(n)$ &
$C_{F3}^{(2)}(n)$ \\   \hline 1& 0 & -4. & 0 &    -52.          \\
2& 0.4444 &    -1.7778  &-3.6395    &    -47.4720
          \\ 3& 3.2222 &     1.6667      &  27.8275 &
          -12.7151                    \\ 4&  6.0667 &   4.8667      &
74.3921 &   37.1171                     \\
5&  8.7259 &    7.7481      & 129.8530 &    95.4086
          \\ 6&  11.1767 &   10.3513     & 190.3464 &
          158.2912                     \\ 7&   13.4368 &   12.7225
& 253.8826 &    223.8978                    \\
8&   15.5299 &  14.9003     &  319.1082 &  290.8840
         \\ 9&   17.4782 &  16.9152     &  385.2570 &
         358.5874                      \\ 10&  19.3006 &   18.7915
& 451.7634 &   426.4422                      \\
11&   21.0130 &   20.5483     & 518.3019 &    494.1881
\\ 12&   22.6284 &   22.2011     &  584.6076 &   561.5591 \\
13& 24.1580 & 23.7624 & 650.5323 & 628.4539 \\ 14& 25.6109 & 
25.2427 & 715.9513 & 694.7397 \\ \hline \end{tabular} \end{center}
{{\bf Table 1.} The numerical values of the NLO and NNLO coefficients
for the coefficient functions of the NS moments of $F_2$ and $xF_3$
SFs at  $f=4$ number of flavours.} \
\vspace{1cm}

\begin{center}
\begin{tabular}{||r|c|c|c||}
\hline
n  & $\gamma_{NS,F_2}^{(1)}(n)$&$\gamma_{NS,F_3}^{(1)}(n)$
& $\gamma_{NS}^{(2)}(n)$ \\
\hline
          1&    2.5576 &       0     &           306.6810           
          \\ 2&   71.3745&      71.2410 &           612.0598  \\ 3&  
          100.8013&      100.7819 &          837.4264  \\ 4&  
          120.1447&      120.1401 &         1005.8235  \\ 5&  
          134.9049&      134.9035 &         1135.9189  \\ 6&  
          147.0029&      147.0024 &         1242.0056  \\ 7&  
          157.3323&      157.3321 &         1334.0017  \\ 8&  
          166.3862&      166.3861 &         1417.4506  \\ 9&  
          174.4683&      174.4682 &         1493.5205  \\ 10&  
         181.7808&      181.7808 &         1559.0048  \\ 11&  
         188.4662&      188.4662 &         ?  \\ 12&  
         194.6293&      194.6293 &          ?\\ 13&  
         200.3496&      200.3496 &         ?  \\ 14&  
         205.6891&      205.6891 &         ?   \\ \hline
\end{tabular} \end{center}
{{\bf Table 2.} The used numerical expressions for the NLO and NNLO 
coefficients of the anomalous dimensions of the moments of   the NS 
SFs at $f=4$ number of flavours.} \vspace{1cm}

\begin{center}
\begin{tabular}{||r|c|c|c||}
\hline
n  &
$\gamma_{NS,F_2}^{(1)}(n)$&$\Delta_{\gamma}$
& $\Delta_{\gamma}(n)/\gamma_{NS,F_2}^{(1)}(n)$ \\
\hline
          2&
          71.3745&      $1.335\times 10^{-1}$ &  $1.87\times 10^{-3}$
           \\
          4&  120.1447&      $4.6\times 10^{-3}$ &
           $3.83\times 10^{-5}$ \\  6&  147.0029&      $5\times
          10^{-4}$ &   $3.4\times 10^{-6}$  \\   8& 166.3862&
          $1\times 10^{-4}$ &   $6.01\times 10^{-7}$  \\
\hline
\end{tabular} \end{center}
{{\bf Table 3.} The related numerical
differences between the NLO coefficients of the anomalous dimensions 
of the NS moments of $F_2$ and $xF_3$ SFs at  $f=4$ number of
flavours. } 
\vspace{1cm}

\begin{center}
\begin{tabular}{|c|c|c||c|c||} \hline
 $Q^2$ cut (GeV$^2$)& $x$ cut & Order &  
 $\Lambda_{\overline{MS}}^{(4)}$ (MeV) 
& $\chi^2$($xF_3$)/nep  \\  
\hline 
$Q^2>5 $ & $x<0.7$     & NLO & $178 \pm 29$ & $81.6/81$ \\ 
& CCFR result &     & $170 \pm 31$ & $83.8/81$\\ 
  $Q^2>10$ & $x<0.7$     & NLO & $182 \pm 30$ & $64.5/65$\\ 
& CCFR result &     & $171 \pm 32$ & $66.4/65$ \\ $Q^2>15$ & $x<0.7$     
         & NLO & $191 \pm 34$  &$51.4/54$ \\ & CCFR result &     & 
$179 \pm 36$ & $53.5/54$ \\ $Q^2>15$ & $F_2~ x>0.5$ & NLO & $235 \pm 
28$ &  $51.9/54$          \\ & CCFR result &     & $210 \pm 28$ & \\  
\hline \end{tabular} \end{center}
{{\bf Table 4.} The comparison of the results of our NLO fits 
of $xF_3$ CCFR data of Ref.[1] with the 
results provided by the CCFR collaboration in Ref.[1] ($\chi^2$ is 
normalized to the number of experimental points (nep).)} \vspace{1cm}

\begin{center}
\begin{tabular}{|c|c|c||c|c||} \hline
 $Q^2$ cut (GeV$^2$)& Order & $N_{max}$ & 
 $\Lambda_{\overline{MS}}^{(4)}$ (MeV) & $\chi^2$($xF_3$)/nep  
\\ \hline  
$Q^2>10 $ & LO    & $10$ & $131 \pm 26$ 
& $76.7/65$ \\ $Q^2>10 $ & NLO   & $10$ & $182 \pm 30$ & $64.5/65$ \\ 
\hline $Q^2>10 $ & NNLO  & $6$  & $184 \pm 31$ & $57.6/65$ \\ 
\hline \end{tabular} \end{center}
{{\bf Table 5.} The results of our  NNLO analysis of the  CCFR data of 
Ref.[1] for $xF_3$ SF.} \vspace{1cm}

\begin{center}
\begin{tabular}{|c|c|c||c|c||} \hline
 $Q^2$ cut (GeV$^2$)& Order & $N_{max}$ & 
 $\Lambda_{\overline{MS}}^{(4)}$ (MeV) & $\chi^2$($xF_3$)/nep   \\ 
\hline 
$Q^2>15 $ & LO    & $10$ & $208 \pm 26$ &  
62.3/54 \\ $Q^2>15 $ & NLO   & $10$ & $235 \pm 28$ &  51.9/54  \\ 
\hline $Q^2>15 $ & NNLO  & $6$  & $210 \pm 24$ &   44.2/54  \\ \hline
\end{tabular} \end{center}
{{\bf Table 6.} The results of our combined NNLO NS
analysis of the  CCFR data of Ref.[1]  for
$xF_3$ and $F_2$ SFs.} \vspace{1cm}

\begin{center}
\begin{tabular}{|c|c|c|c|c|} \hline
  Order& & $\alpha_s (M_Z)$ & $\Delta \alpha_s^{stat}$
 & $\Delta \alpha_s^{sys}$  \\
\hline
  NLO   & $xF_3$     & 0.108 &  0.003& 0.005 \\
  NLO   & $xF_3+F_2$ & 0.112 &  0.002& 0.003 \\
  NNLO  & $xF_3$     & 0.109 &  0.003& 0.005 \\
  NNLO  & $xF_3+F_2$ & 0.111 &  0.002& 0.003 \\
\hline
\end{tabular}
\end{center}
{{\bf Table 7.} The obtained  NLO and NNLO results for
$\alpha_s(M_Z)$.}

\end{document}